\documentclass[aps,pre,twocolumn]{revtex4}
\usepackage{amsmath,bm,epsfig,,subfigure}
\usepackage{color}
\usepackage[normalem]{ulem}

\def\Fbox#1{\vskip1ex\hbox to 8.5cm{\hfil\fboxsep0.3cm\fbox{%
  \parbox{8.0cm}{#1}}\hfil}\vskip1ex\noindent}  

\newcommand{\sFrac}[2]{{\textstyle\frac{#1}{#2}}}
\newcommand{\C}[1]{{\mathcal{#1}}}    
\let \= \equiv \let\*\cdot \let\~\widetilde \let\-\overline

\let \= \equiv \let\*\cdot \let\~\widetilde \let\^\widehat \let\-\overline

\begin{document}
\title{Propagation Mechanism of Brittle Cracks}
\author{Itamar Procaccia and Jacques Zylberg}
\affiliation{Dept. of Chemical Physics, The Weizmann Institute of
Science, Rehovot 76100, Israel}
\date{\today}
\begin{abstract}
We employ a recently developed model that allows the study of two-dimensional brittle crack propagation under fixed grip boundary conditions. The crack development highlights the importance of voids which
appear ahead of the crack as observed in experiments on the nano-scale. The appearance of these voids
is responsible for roughening the crack path {\em on small scales}, in agreement with theoretical expectations. With increasing speed of propagation one observes the branching instabilities that were reported in experiments. The simulations allow understanding the phenomena by analyzing the elastic stress field that accompanies the crack dynamics.
\end{abstract}
\maketitle

\section{Introduction}

In laboratory experiments one studies crack propagation in elastic media by holding a stretched
slab of material with a grip, and initiating the crack by making a small notch at one end of the sample
\cite{99HHMS}. When the notch exceeds the Griffith's length \cite{21Gri}, the crack propagates with increasing speed until there is a balance between the release of elastic energy and the creation of surface energy \cite{98Fre,99MF}. It turned out that molecular dynamics simulations failed systematically to mimic this set up, forcing simulators to pull continuously on the boundaries to achieve a propagating crack \cite{11HKL}. With fixed grip boundary conditions
crack tended to slow down and stop propagating. The inability to simulate a brittle crack by molecular dynamics gave rise to claims that it would be impossible to advance a brittle crack without continuous stretching.  The fundamental reason for this long-standing problem was understood recently. In Ref. \cite{11DKPZ} it was shown that the solution of this conundrum  lies in the range of the inter-particle
potential. By changing the potential range one can go from a brittle to a ductile crack, and the latter is stopped by plastic dissipation. Only a brittle crack can support itself with a fixed grip. Brittleness was shown to be guaranteed by choosing a potential range that is of the order of the inter-particle distance. In this paper we present results of molecular dynamics simulations of brittle cracks under a fixed grip that are based on this understanding.

One of the interesting issues that can be studied by such simulations is the proposed
existence of voids that open up before the crack, acting as pointers for further propagation. This issue had been quite controversial. While some experiments, and in particular those pertaining to ultra-slow crack propagation in glass in wet atmosphere, strongly indicated the appearance of
voids ahead of the crack \cite{Bou}, other experiments failed to observe such voids, and claimed that they are irrelevant to the propagation of brittle cracks. In a theoretical study, Ref. \cite{04BMP} constructed a model of
void-dominated crack propagation and attempted to explain the observed roughness of brittle cracks
by the randomness associated with the positioning of the voids ahead of the crack. The numerical simulations shown below appear to vindicate this approach almost verbatim. Voids do appear, and their appearance is random as stipulated in Ref. \cite{04BMP}, and see Sect. \ref{voids} for more details.

Another issue of interest is the instabilities of the propagating crack when its velocity increases.
The simulations show very clearly branching instabilities with subsequent competition between the
two branches that typically result in the death of one branch in favor of the other.

In Sect. \ref{protocol} we present the details of the preparation of the samples for the
molecular dynamics simulations. In Sect. \ref{Grif} we show that our cracks are brittle by
demonstrating that they start running precisely when the Griffith's criterion met. In Sect. \ref{rough} we discuss the appearance of voids
ahead of the crack and the resulting roughening of the crack path. Finally, in Sect. \ref{inst} we present a brief discussion of the observed instabilities. The last section presents a short summary and indicates the road ahead.

\section{Molecular Dynamics}
\label{protocol}

\begin{figure}[h]
\centering
\hskip -1. cm
\includegraphics[scale = 0.65]{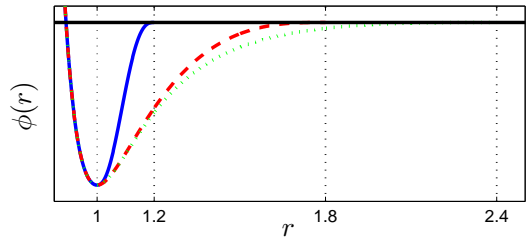}
\caption{Color online: Examples of three different potentials with $r_{\rm co}= 1.2, 1.8$ and 2.4 generated using Eq. (\ref{potential}). We used the potential with $r_{\rm co}=1.2$ for the purpose of this paper.}
\label{potentials}
\end{figure}

For the numerical experiments we employ a generic glass former in 2-dimensions in the form of a 50-50 binary mixture
of `small' and `large' particles, chosen to avoid any crystallization. In fact the particles interact
by inter-particle potentials as shown in Fig. \ref{potentials}, with the analytic form
\begin{widetext}
\begin{equation}\label{potential}
\phi\!\left(\!\sFrac{r_{ij}}{\lambda_{ij}}\!\right) =
\left\{\begin{array}{ccl}
&&4\varepsilon\left[\left(\frac{\lambda_{ij}}{ r_{ij}}\right)^{12}-\left(\frac{\lambda_{ij}}{r_{ij}}\right)^{6}\right]
\ , \hskip 3.5 cm\frac{r_{ij}}{\lambda_{ij}}\le \frac{r_{\rm min}}{\lambda}\\
&&\varepsilon\left[a\left(\frac{\lambda_{ij}}{ r_{ij}}\right)^{12}-b\left(\frac{\lambda_{ij}}{ r_{ij}}\right)^{6}
+ \displaystyle{\sum_{\ell=0}^{3}}c_{2\ell}\left(\sFrac{r_{ij}}{\lambda_{ij}}\right)^{2\ell}\right]
  \ , \hskip 0.5 cm\frac{r_{\rm min}}{\lambda}<\frac{r_{ij}}{\lambda_{ij}}<\frac{r_{\rm co}}{\lambda} \\
&&\quad\quad\quad\quad 0
   \ , \hskip 5.4 cm  \frac{r_{ij}}{\lambda_{ij}} \ge\frac{r_{\rm co}}{\lambda}
\end{array}
\right.\!\!
\end{equation}
\end{widetext}
Here $r_{\rm min}/\lambda_{ij}$ is the length where the potential attain it's minimum, and $r_{\rm co}/\lambda_{ij}$
is the cut-off length for which the potential vanishes. The coefficients $a,~b$ and $c_{2\ell}$ are chosen such
that the repulsive and attractive parts of the potential are continuous with two derivatives at the potential minimum and the potential goes
to zero continuously at $r_{\rm co}/\lambda_{ij}$ with two continuous derivatives as well.
The interaction length-scale $\lambda_{ij}$ between any two particles $i$ and $j$ is $\lambda_{ij} = 1.0\lambda$, $\lambda_{ij} = 1.18\lambda$
and $\lambda_{ij} = 1.4\lambda$ for two `small' particles, one `large' and one `small' particle and two `large' particle respectively. The
unit of length $\lambda$ is set to be the interaction length scale of two small particles, $\varepsilon$ is the unit of energy and $k_B = 1$.
\begin{figure}[h]
\centering
\includegraphics[scale = 1.00, angle = 90]{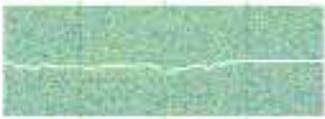}
\caption{Color Online: a typical crack in a system of size 1000$\lambda\times$3000$\lambda$ which is self sustaining under fixed grip conditions with $\gamma=2.5\%$. The initial cut was of length $L=750\lambda$. Note
the roughening on small scales which is discussed in Sect. \ref{rough}.}
\label{crack}
\end{figure}

We used a modified Berendsen thermostat which couples a constant number of particles to the bath, regardless of the system size \cite{10KLPZ}. The preparation protocol starts with equilibrating the sample at high temperature and pressure, using periodic boundary conditions. The systems are then cooled to temperature $T=10^{-2}$ while keeping the pressure high, in order to avoid the creation of any holes in the material. Then the pressure is reduced to zero, $P=0.0$, such that the periodic boundary conditions could be removed;  particles forming the right and left walls were frozen, but the upper and lower boundaries were rendered free. The brittle
crack experiment starts by loading the system uniaxially (with a constant velocity such that $v_{\rm wall} \ll c_s/10$) until a desired stress
is reached, and then the side walls are held fixed. A cut is then implemented by the cancelation of forces crossing an imaginary line of desired length which starts at the lower boundary. The evolution of the crack is simulated by molecular dynamics coupled to a heat bath at $T = 0.0$ and requires no further loading of the system. The results presented below were created using two different geometries, one
of width 1000$\lambda$ and length 3000$\lambda$, and the other being a square sample of 1600$\lambda\times 1600\lambda$. The density is the zero temperature and zero pressure density $\rho=0.745$.

In Fig. \ref{crack} we present a typical crack that results from this procedure, in which the loading $\gamma$ was
$\gamma=2.5\%$ and an initial crack of length 750$\lambda$. One observes the typical roughness that
is discussed below in Sect. \ref{rough}.

\section{The Griffith Length}
\label{Grif}

Upon making the initial cut, the system always has a microscopic plastic response which
however stops if the length of the cut is smaller than the Griffith length. To demonstrate that
our running cracks are indeed brittle we test the Griffith's criterion in our simulation.

The length of the initial cut that evolves spontaneously into a crack was determined by Griffith \cite{21Gri}, comparing the energy release from the stress field to the energy consumed by the creation of two new fresh surfaces by the advancing crack.  With the stress exerted on the slab being $\sigma$, the critical length $L_c$ is determined by the bulk modulus $E$ and the surface energy per unit length $\epsilon$ via the relation
\begin{equation}
\sigma \sqrt {L_c }= \sqrt{\frac{2E\epsilon}{\pi}} .
\label{griff}
\end{equation}
 In  our simulations we are able to measure the critical length $L_c$ for a given loaded system. Since all the other material parameters appearing in Eq. \ref{griff} are measured independently, we can show that our model crack propagation agrees with the expected physics of brittle cracks.

\begin{figure}
\includegraphics[scale = 0.65]{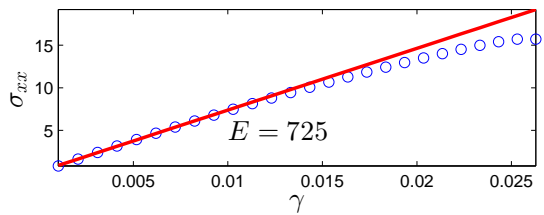}
\caption{Color Online: A typical stress-strain curve measured from a uniaxial straining experiment. We read the Bulk modulus $E\approx 725$ from the slope and evaluate the stress for the theoretical estimates of the Griffith's Length.}
\label{bulkModulus}
\end{figure}

 The stress field $\sigma$ is simply determined by our grip boundary conditions. The bulk modulus $E$ is read from the linear part of a uniaxial straining experiment shown in Fig. \ref{bulkModulus}.
To estimate the surface energy in our system we computed the total energy before and after making
the initial cut of length $L$. Taking the difference and dividing by $L$ we get the estimate
\begin{figure}
\includegraphics[scale = 0.75]{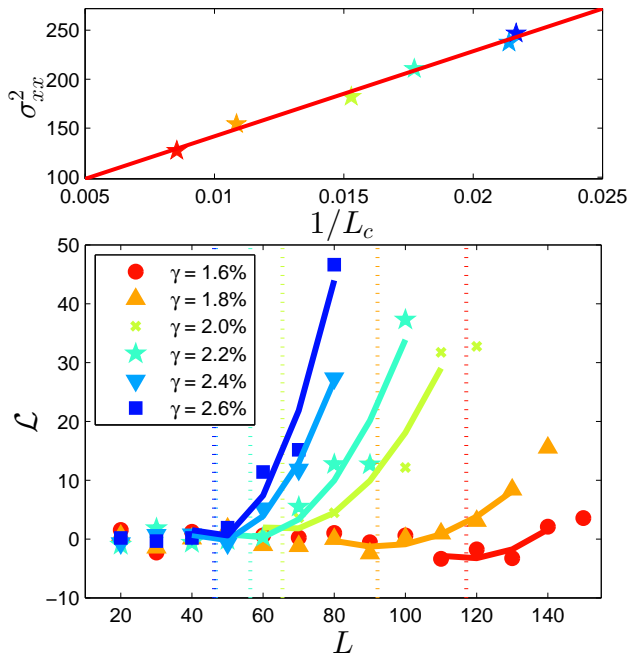}
\caption{Color Online: Upper Panel: The linear relationship confirms the compliance of our model to Griffith's Law. The slope corresponds to the theoretical estimation emanating from Eq. (\ref{griff}).
Lower Panel: Plotted are the average values of the crack length ${\C L}$ within 30 time units of the experiment. We read the critial Griffth's length $L_c$ at the size of L for which ${\C L}>0$.}
\label{Griff}
\end{figure}
\begin{equation}
\epsilon \approx 15 \ .
\end{equation}

At this point we employed 50 independent samples that were equilibrated at high temperature and quenched to the glassy state. The glassy samples were then loaded quasi-statically to strain values in the
range $\gamma= 1.6, 1.8, 2.0, 2.2, 2.4, 2.6$. In order to probe the Griffith's Length $L_c$ we tracked the behavior of our sample during the first 30 time units of our simulation to see whether a crack has started to run as a result of the introduced cut. For each sample we repeated this procedure for various values of $L$. The results, (averaged over 50 samples) are presented in the bottom panel of Fig. \ref{Griff}. We show the length of the evolving crack, denoted as $\C L$, as a function the initial
cut size $L$. Where the length $\C L$ begins to run determines $L_c$ for the conditions at hand.

In the upper panel of the same figure we plot  $\sigma^2$ vs. $1/L_c$. The resulting linear plot is in
agreement with Eq. (\ref{griff}), with the slope being close to the theoretical expectation of $2E\epsilon/\pi$.

\section{Roughening resulting from voids ahead of the crack}
\label{rough}

The present numerical simulations allows the verification of the role of voids that form
ahead of the crack, determining, at least on short length scales, the path chosen by the
crack as it develops in the amorphous solid. We note that an amorphous solids is not an ideal
elastic material, in which mathematically straight cracks are possible. The randomness of the
material must show up in the geometry of the crack in this way or another. The voids ahead of the
crack serve as pointers for the forthcoming propagation of the crack, and as is shown below,
their random positions determines the roughening of the crack path.

\begin{figure}
\includegraphics[scale = 0.50]{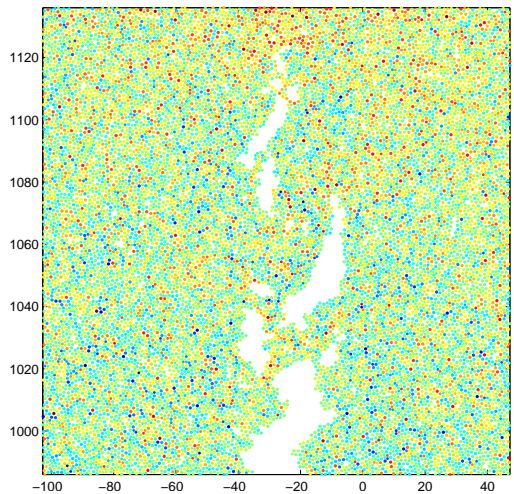}
\caption{Color Online: A typical crack tip with the voids that opened ahead of the crack}
\label{voids}
\end{figure}

\begin{figure}
\includegraphics[scale = 0.7]{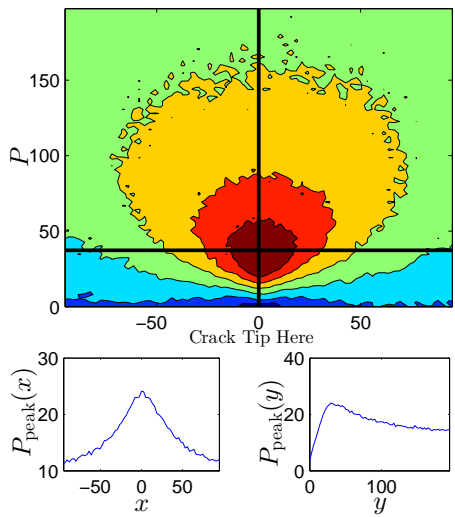}
\caption{Color Online: Upper Paner: The pressure field in a fixed frame tracking the crack tip, averaged over time. The color code is brown for lowest pressure and blue the highest. The two black lines are the horizontal and vertical cross section of this measured pressure field and are shown in the bottom left and right panels respectively. This last picture demonstrates the fact that the maximal pressure lies ahead of the crack tip where voids are being created.}
\label{pressure}
\end{figure}

In Fig. \ref{voids} we show a typical crack tip with the voids that opened ahead.
The physical reason why voids may prefer to open ahead of the crack and not on the crack tip was explained in Ref. \cite{04BMP}. The argument is based on the fact that voids open due to plastic
yield, and they do this where the pressure is maximal. Near the crack tip there is a process zone where the pressure is increasing going outward, until one hits the maximal pressure curve which connects
with the outer elastic solution, see Fig. \ref{pressure}. In this figure we show the average pressure as measured in the numerical simulation in a fixed window tracking in time the crack tip.
In particular one should pay attention to the lower right panel in Fig. \ref{pressure} which shows that
the maximal pressure lies ahead of the tip, at a distance $\xi$ from the tip. This figure should be compared with Fig.2 of Ref. \cite{04BMP} which it directly vindicates.

\begin{figure}
\includegraphics[scale = 0.70]{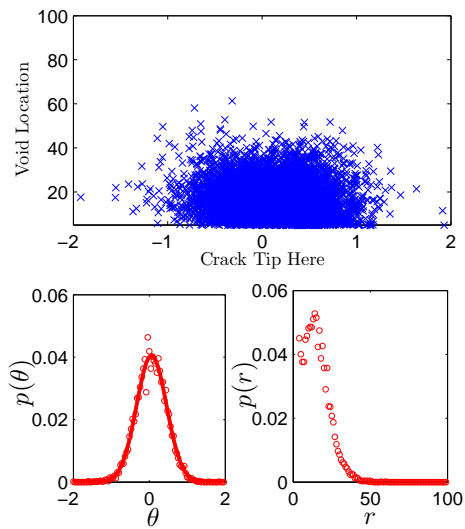}
\caption{Color Online: Upper panel: the actual positions of the voids appearing ahead of the
crack during the time evolution of the crack. These results are histogrammed in the lower two panels
to provide the angular distribution of the void positions and the radial distribution in the forward
direction.}
\label{distribution}
\end{figure}

Obviously, the void that opens ahead of the tip can have a broadly distributed position
around the forward direction (where the probability to form a void is maximal). An actual measurement
of this distribution for the present simulation is shown in Fig. \ref{distribution}.
Shown are the actual positions of the voids as they appear ahead of the crack during the time
evolution, and in the lower panels the probability distribution for the void to appear at an
angle $\theta$ with respect to the forward direction, and at a distance $r$ along the forward
direction. These data should be compared with Fig. 3 of Ref. \cite{04BMP}.

Two points are worth stressing: first, the angular distribution of void positions will be the source of roughening of the crack - the probability to fall along the forward direction is not high enough
compared to positive or negative angles with the respect to the forward direction. Second, once
a void appears on, say, a positive angle, the next void will have an even higher probability for a positive angle, meaning that the rough crack is expected to show a roughening exponent higher than 1/2. Indeed, such a persistent random walk is always expected to show exponents higher than
1/2, whereas anti-persistent random walks are characterized by a roughening exponent smaller than 1/2.

\begin{figure}
\includegraphics[scale = 0.60]{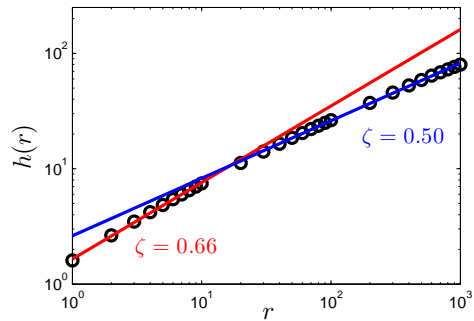}
\caption{Color online: The min-max methodology shows the persistent nature of the crack meandering for small values of $r$ with a roughness exponent of $\zeta=0.66$. For larger values of $r$ the roughness exponent is reduced to $\zeta=0.50$ and the crack meandering loses its persistence.}
\label{roughness}
\end{figure}

The roughening exponent was measure here in the usual way, i.e. considering the crack as a graph
$y(x)$ and determining $h(r)$ according to
\begin{equation}
h(r) \equiv \langle {\rm max}\{y(\tilde x\}\}_{x<\tilde x<x+r} - {\rm min}\{y(\tilde x\}_{x<\tilde x<x+r}\rangle_x \ .
\end{equation}
For a self-affine graph the scaling exponent $\zeta$ is defined via the scaling relation
\begin{equation}
h(r) \sim r^\zeta \ .
\end{equation}
We show this quantity in a log-log plot in Fig. \ref{roughness}. As expected the function exhibits
a persistent scaling exponent of $\zeta\approx 0.66$ for scales $r<30$, and then a cross over
to a random graph without persistence or anti-persistence for higher scales.

\section{Instabilities}
\label{inst}
\begin{figure}
\includegraphics[scale = 0.60]{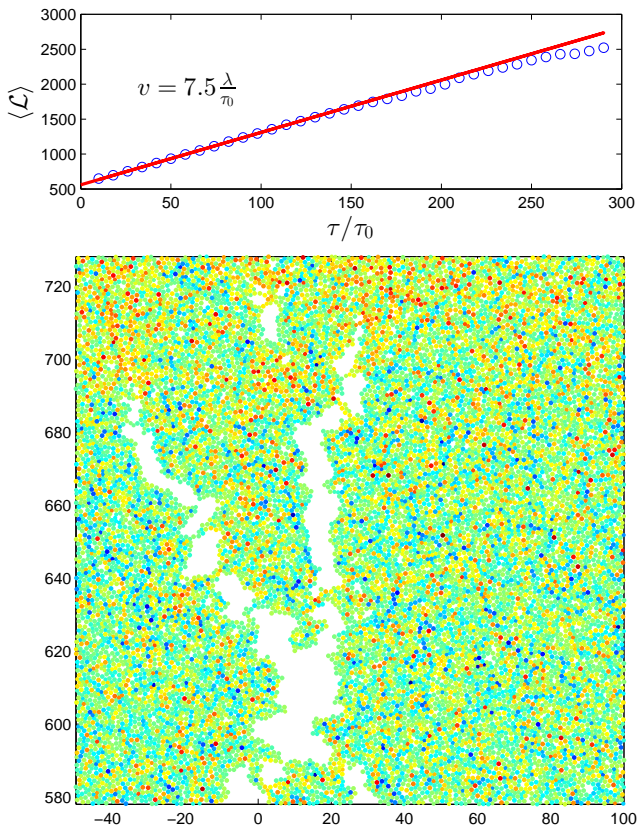}
\caption{Color online: Upper panel: The crack length as a function of time; the velocity $v$ is read from the slope. Lower panel:  A typical branching event. One sees the two crack tips advancing simultaneously for a while, until the competition between them selects one or the other. Here the right crack wins, as is evidenced by the void ahead of it.}
\label{branch}
\end{figure}

Crack dynamics are governed by the balance of energy at the crack tip. The influx of elastic energy through the crack-tip is used to create surface energy behind the advancing tip. With increasing crack velocity there is not sufficient surface to store the energy that is released. Therefore the system resorts to instabilities in the form of branching and oscillations in order to increase the amount of surface created per unit length on the axis of propagation. In our simulations we find that when the velocity of the crack tip reaches about  $30\%$ of the Rayleigh speed, one begins to observe crack branching. We note that our system never develops two independently propagating cracks, but rather reduced the velocity of propagation through attempted branchings. Oscillations were not observed.
A typical branching event is shown in Fig. \ref{branch}, where we see the two crack tips as they
still grow simultaneously. The competition between them always results in the demise of one of
the growing cracks in favor of the other, until the next branching event.

\section{summary and discussion}
In summary, we have demonstrated that molecular dynamics can be usefully employed to study brittle
crack propagation by selecting an appropriate interparticle potential. The resulting cracks are growing by nucleating voids ahead of the crack tip in much the same way that was anticipated by Ref. \cite{Bou}
and theorized in Ref. \cite{04BMP}. This mechanism of growth is responsible for roughening of the
crack path on small scales, but as long as side branching does not commence, the crack is globally flat on macroscopic scales, as expected theoretically \cite{06BBP}.

In future work molecular dynamics simulations will be employed in three dimensions where the issue
of micro-branching \cite{99MF} can be studied and compared with experimental results.

\acknowledgments
We thank Laurent Bou\'e for useful discussion. This work had been supported in part by the German Israeli Foundation, the Israel Science Foundation
and the European Research Council under an "ideas" grant STANPAS.

\end{document}